\documentclass[10pt,twoside]{article}

\usepackage{fancyhdr}
\usepackage{titlesec}
\usepackage{cite}
\usepackage{ifthen}
\usepackage{graphicx}
\usepackage{float}
\usepackage{amsmath,amssymb}

\titleformat{\section}{\centering\large\bfseries}{\S\arabic{section}}{1em}{}
\newboolean{first}
\setboolean{first}{true}

\begin{document}

\setlength\abovedisplayskip{2pt}
\setlength\abovedisplayshortskip{0pt}
\setlength\belowdisplayskip{2pt}
\setlength\belowdisplayshortskip{0pt}

\title{\bf \Large  Multiple routes transmitted epidemics on multiplex networks
\author{Dawei Zhao$^{a,b}$   \  \ Lixiang Li$^{a,b}$   \  \ Haipeng Peng$^{a,b}$   \  \ Qun Luo$^{a,b}$\ \ Yixian Yang$^{a,b}$ \\ \small \it $^{a}$Information Security Center, Beijing University of Posts and
Telecommunications, \\
\small \it Beijing 100876, China. \\ \small \it $^{b}$National
Engineering Laboratory for Disaster Backup and
Recovery, \\
\small \it Beijing University of Posts and Telecommunications,
Beijing 100876, China. }\date{}} \maketitle

\footnote{E-mail address: dwzhao@ymail.com (Dawei Zhao)}

\begin{center}
\begin{minipage}{135mm}
{\bf \small Abstract}.\hskip 2mm {\small This letter investigates
the multiple routes transmitted epidemic process on multiplex
networks. We propose detailed theoretical analysis that allows us to
accurately calculate the epidemic threshold and outbreak size. It is
found that the epidemic can spread across the multiplex network even
if all the network layers are well below their respective epidemic
thresholds. Strong positive degree-degree correlation of nodes in
multiplex network could lead to a much lower epidemic threshold and
a relatively smaller outbreak size. However, the average similarity
of neighbors from different layers of nodes has no obvious effect on
the epidemic threshold and outbreak size.}
\end{minipage}\end{center}
\begin{center}
\begin{minipage}{135mm}
\end{minipage}
\end{center}

\section{Introduction}
\label{}

In recent years, various types of epidemics have occurred frequently
and spread around the world, causing not only great economic losses,
but also widespread public alarms. For example, the intense outbreak
of SARS caused 8,098 reported cases and 774 deaths. Within weeks,
SARS spread from Hong Kong to infect individuals in 37 countries in
early 2003 [1]. An outbreak of mobile viruses occurred in China in
2010. The `Zombie' virus attacked more than 1 million smart phones,
and created a loss of \$300,000 per day [2]. And we have also
witnessed how social networks being used for citizens to share
information and gain international support in the Arab Spring [3].
In view of these situations, it is thus urgent and essential to have
a better understanding of epidemic process, and to design effective
and efficient mechanisms for the restraint or acceleration of
epidemic spreading.

Valid epidemic spreading models can be used to estimate the scale of
an epidemic outbreak before it actually occurs in reality and
evaluate new and/or improved countermeasures for the restraint or
acceleration of epidemic spreading. In the last decade, there have
been extensive studies on the modeling of epidemic dynamics [4-10],
and various protection strategies have been proposed and evaluated
[11-18]. However, these existing researches have been dominantly
focusing on the cases that epidemics spread through only single
transmission route. While in reality, many epidemics can spread
through multiple transmission routes [19] simultaneously. For
example, it has been well recognized that AIDS can propagate via
three routes simultaneously including sexual activity, blood and
breast milk; rumor or information can spread among human through
verbal communication and social networks; malwares can move to
computers by P2P file share, email, random-scanning and instant
messenger [20]; and some mobile malwares can attack smart phones
through both short messaging service (SMS) and bluetooth (BT) at the
same time [21]. In this letter, the epidemic which spreads via
single transmission route and multiple transmission routes are
called single route transmitted epidemic and multiple routes
transmitted epidemic, respectively. When a multiple routes
transmitted epidemic spreading on a network, the network node could
be infected via one of the transmission routes even if it cannot be
infected via the other routes. And the node can be infected with a
higher probability if it can be infected via more than one
transmission route of the epidemic. Therefore, the range and the
intensity of the multiple routes transmitted epidemic will be
greater than those of the traditional single route transmitted
epidemic. Meanwhile, different transmission routes are supported by
different networks. For instance, the underlying network of the
mobile malware which propagates via SMS is a SMS network formed
based on the social relationships among mobile users. And the BT
network formed according to the geographically positions of mobile
devices is the underlying network of the mobile malware which can
spread through BT. Therefore, the underlying network of the multiple
routes transmitted epidemic is actually a multiplex network [22-25],
rather than a single network. Multiplex network can be regarded as a
set of coupled layered networks in which each layer could have very
particular features different from the rest and support different
dynamical processes. Based on the above analyses, the study of
multiple routes transmitted epidemic on multiplex network is
definitely a very meaningful and necessary topic.

To the best of our knowledge, the theory describing the multiple
routes transmitted epidemic process on multiplex network has not
been fully developed yet. In this letter, we propose and evaluate a
two routes transmitted epidemic spreading on multiplex network with
two network layers following the typical
Susceptible-Infected-Removed (SIR) model [6,7]. But the proposed
research methods can be easily extended to analyze the epidemics
which spread via any number of transmission routes. By mapping the
SIR model into the bond percolation [7], we develop equations which
allow accurate calculations of epidemic threshold [6] of the
multiplex network and outbreak size [6] of the epidemic. It is found
that the epidemic can spread across the multiplex network even if
the two network layers are well below their respective epidemic
thresholds. We also introduce two quantities for measuring the level
of inter-similarity between these two layers. One is the average
similarity of neighbors (ASN) from different layers of nodes. ASN
evaluates how many neighbors of nodes in one layer are also their
neighbors in another layer. We find that both epidemic threshold and
outbreak size are not significantly affected by the ASN. The second
quantity is the degree-degree correlation (DDC) of nodes which
describes the correlation of nodes' degrees in one layer and that in
another layer. Positive DDC indicates that high degree nodes in one
layer are also high degree ones in another layer, and vise versa. It
is found that strong positive DDC could lead to a clearly lower
epidemic threshold and a relatively smaller outbreak size.

Some symbols used throughout this letter and their meanings are
summarized in Table 1.

\begin{table}\centering{
\caption{Symbols used in this letter and their
meanings.}\footnotesize \label{tab:1}
\begin{tabular}{lllll}
\hline\noalign{\smallskip} Symbols & Meanings\\
\hline\noalign{\smallskip}
ASN & Average similarity of neighbors from different layers of nodes.\\
DDC & Correlation of nodes' degrees in one layer and that in another layer.\\
$\lambda_A$ & The probability that a susceptible node is infected
only via route-$A$.\\
$\lambda_B$ & The probability that a susceptible node is infected
only via route-$B$.\\
$\lambda_C$ & The probability that a susceptible node is infected
via route-$A$ and\\
& route-$B$ simultaneously.\\
$(\lambda_A, \lambda_B)$ & The spreading rate of a two routes
transmitted epidemic, where $\lambda_A$\\
& and $\lambda_B$ are the spreading rates of this epidemic when
spreading on\\
& layer-$A$ and layer-$B$, respectively.\\
$k_A$ & Degree of node in layer-$A$.\\
$k_B$ & Degree of node in layer-$B$.\\
$k_C$ & The number of same neighbors of node in layer-$A$ and
layer-$B$.\\
$k_M$ & Vector degree of node on multiplex network.\\
$\{(\lambda_A, \lambda_B)_c\}$ & Epidemic threshold of multiplex network with two layers.\\
$s$ & Outbreak size of epidemic.\\
\hline
\end{tabular}}
\end{table}

\section{Models and analysis}
\label{}

\subsection{Multiple routes transmitted epidemic spreading model}
The epidemic spreading model adopted here is the
Susceptible-Infected-Removed (SIR) model which is the most basic and
well-studied epidemic spreading model [6,7]. In the SIR model, the
nodes of the network can be divided into three compartments,
including susceptibles (S, those who are prone to be infected),
infectious (I, those who have been infected), and recovered (R,
those who have recovered from the disease). At each time step, a
susceptible node becomes infected with probability $\lambda$ if it
is directly connected to a infected node. The parameter $\lambda$ is
called the spreading rate. Meanwhile, an infected node becomes a
recovered node with probability $\delta$.

In this letter, what we study is a simple case that a two routes
transmitted epidemic spreads among network individuals. Therefore,
we need to specify the corresponding epidemic spreading processes
separately. It is assumed that these two transmission routes of the
epidemic are route-$A$ and route-$B$, respectively. Then we assume
that a susceptible node becomes infected with probability
$\lambda_A$ or $\lambda_B$ if it can be infected only through
route-$A$ or route-$B$. Besides, if a susceptible node can be
infected via route-$A$ and route-$B$ simultaneously, the probability
that this susceptible node becomes infected is assumed to be
$\lambda_C$. Obviously, $\lambda_C=1-(1-\lambda_A)(1-\lambda_B)$.
Meanwhile, an infected node becomes a recovered node with
probability $\delta$. Without loss of generality, we let $\delta=1$.

\subsection{Multiplex networks model}
Since different transmission routes are supported by different
networks, the underlying network of a two routes transmitted
epidemic should be a multiplex network with two network layers. In
this section, as shown in Fig.1(a), we propose a multiplex network
model which contains two network layers, i.e., layer-$A$ and
layer-$B$. Nodes are the same in both layers, and layer-$A$ and
layer-$B$ are the underlying networks of the epidemic spreading via
route-$A$ and route-$B$, respectively. Fig.1(b) shows this multiplex
network in the form of the superposition of layer-$A$ and layer-$B$.
In the rest of this paper, the multiplex network, unless otherwise
noted, is assumed to be the network in the form of the superposition
of layer-$A$ and layer-$B$.

\begin{figure}[h]
\centering{\includegraphics[scale=0.5,trim=0 0 0 0]{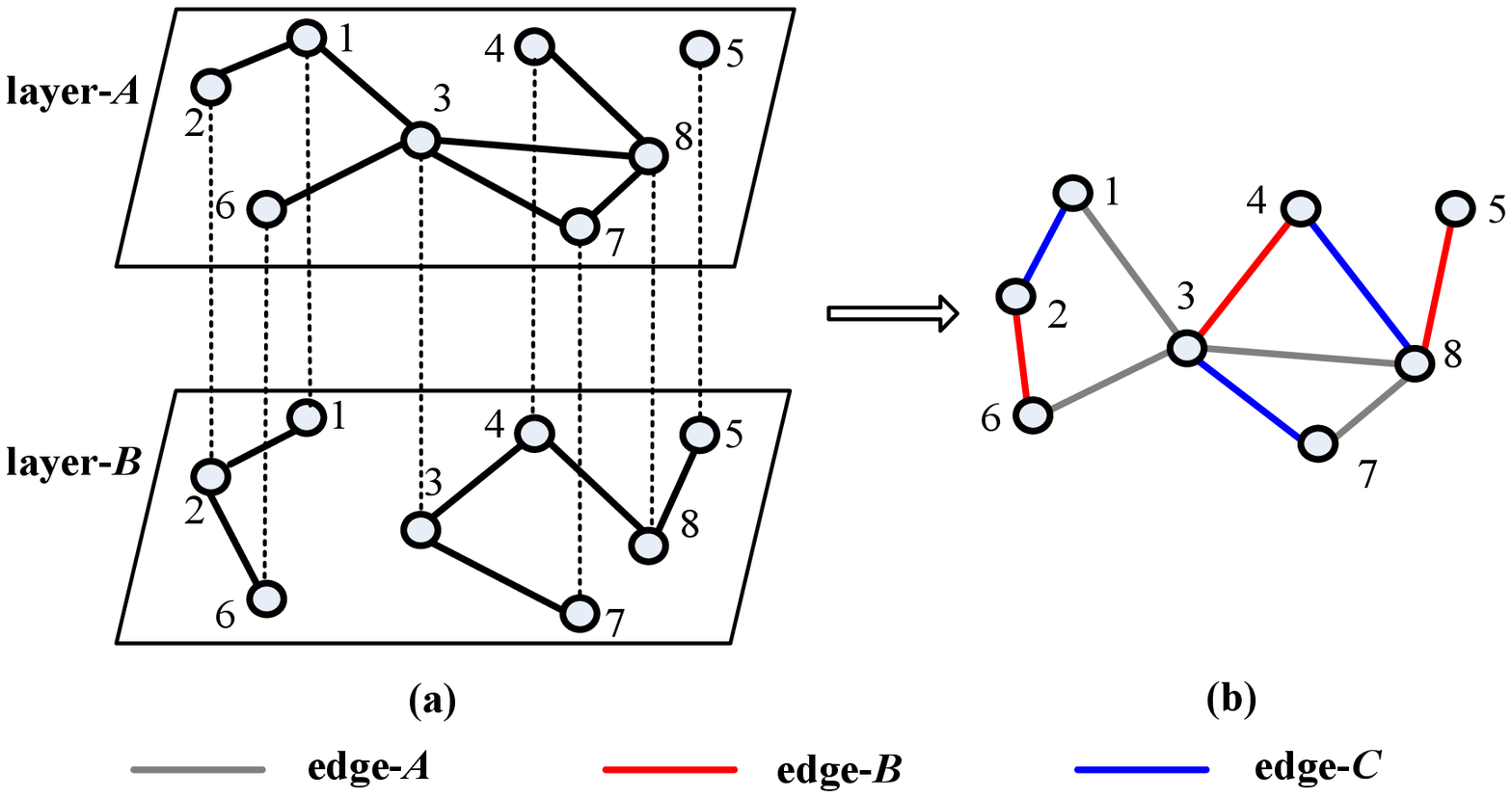}}
\caption{(Color online) (a) A multiplex network with two network
layers, i.e., layer-$A$ and layer-$B$. (b) The multiplex network in
the form of the superposition of layer-$A$ and layer-$B$.}
\end{figure}

Each node in the proposed multiplex network has up to three types of
edges where edge-$A$ belongs only to layer-$A$, edge-$B$ belongs
only to layer-$B$, and edge-$C$ belongs to both layer-$A$ and
layer-$B$. The vector degree $k_M\equiv(k_A- k_C, k_B- k_C, k_C)$ is
used to characterize the node of multiplex network, where $k_A-
k_C$, $k_B- k_C$ and $ k_C$ represent the numbers of edge-$A$,
edge-$B$ and edge-$C$ of the node, respectively. The numerical value
of vector degree of the node is defined by $|k_M|= k_A+k_B- k_C $.
For instance, the vector degree of node 3 in Fig.1 is
$k_M\equiv(4-1, 2-1, 1)$ and its numerical value is 5.

Actually, for the vector degree of node in multiplex network, $k_C$
evaluates how many neighbors of the nodes in layer-$A$ are also
their neighbors in layer-$B$ which can affect the topology of the
multiplex network. Here, we develop a measure, $\alpha$, to assess
the average similarity of the neighbors (ASN) from different layers
of nodes in the multiplex network and it is defined as
\begin{equation}\alpha=\frac{\sum\limits_{i}{k_C}(i)}
{\sum\limits_{i}{|k_M}(i)|},\end{equation} where ${k_C}(i)$ and
$|k_M(i)|$ are the values of $k_C$ and $|k_M|$ of node $i$ of the
multiplex network, respectively. For increasing values of $\alpha$,
more of the neighbors of nodes in one layer are also their neighbors
in another layer and these two layers become more similar. For
$\alpha=1$, these two layers must be identical.

For the node of the multiplex network, it may be a high degree node
in layer-$A$ and a low degree one in layer-$B$, or a high degree
node in layer-$A$ and also a high degree one in layer-$B$. The
influence of the correlation of nodes' degrees in one layer and that
in another layer for the multiple routes transmitted epidemic
dynamics is one of our main research problems. Analogously to the
degree correlation in the single network [26,27] and the network
assortativity in interconnected networks [10], we define the
degree-degree correlation (DDC) of the nodes in multiplex network as
follows
\begin{equation}\beta =\frac{\sum\limits_{k_A}\sum\limits_{k_B}(k_A k_B(p(k_A, k_B)-
(\sum\limits_{k_A}p(k_A, k_B))( \sum\limits_{k_B}p(k_A,
k_B))))}{\sum\limits_{k_B}k_B^2 \sum\limits_{k_A} p(k_A,
k_B)-(\sum\limits_{k_B} k_B \sum\limits_{k_A} p(k_A,
k_B))^2},\end{equation} where $p(k_A, k_B)$ denotes the probability
that a randomly chosen node of multiplex network has degree $k_A$ in
layer-$A$ and $k_B$ in layer-$B$. The two layers are said to be
negative correlation if $\beta < 0$, positive correlation if $\beta
> 0$, and uncorrelation if $\beta = 0$.

\subsection{Calculations of epidemic threshold}

The traditional epidemic threshold [6] of the single network is a
value, $\lambda_c$, above which the epidemic will spread to the
whole network, i.e., the infected nodes will form into a giant
component. Otherwise, the epidemic outbreak will not affect a finite
portion of the nodes and will die out in a finite time. However,
unlike the epidemic threshold of the single network, the epidemic
threshold of the multiplex network with $M$ layers should be a set
of $M$-dimensional points, $\{(\lambda_1,
\lambda_2,...,\lambda_M)_c\}$. The multiplex network with two layers
is taken as an example. As shown in Fig.2, the epidemic threshold of
the two-layers network is a set of 2-dimensional points,
$\{(\lambda_A, \lambda_B)_c\}$. For a two routes transmitted
epidemic with spreading rate $(\lambda_A, \lambda_B)$, where
$\lambda_A$ and $\lambda_B$ are the spreading rates of this epidemic
when spreading on layer-$A$ and layer-$B$ respectively, if
$(\lambda_A, \lambda_B)$ is a point in the grey shaded area of
Fig.2, this two routes transmitted epidemic can spread across the
multiplex network. Otherwise, it will not affect a finite portion of
the nodes and will die out in a finite time.

\begin{figure}[h]
\centering{\includegraphics[scale=0.5,trim=0 0 0 0]{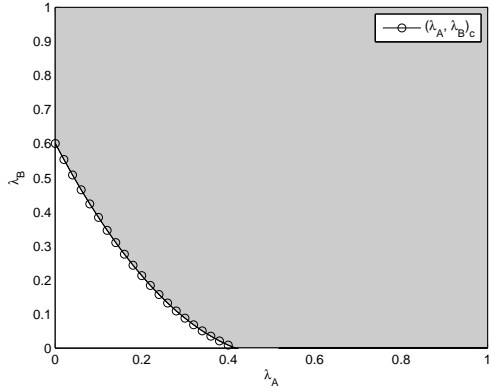}}
\caption{(Color online) The solid line connected with circles
indicates the epidemic threshold of a two layers networks whose two
layers are an ER network comprised of 2000 nodes with average degree
2.858 and an ER network comprised of 2000 nodes with average degree
1.891, respectively. If the spreading rate $(\lambda_A, \lambda_B)$
of a two routes transmitted epidemic is a point belonging to the
grey shaded area, this epidemic can spread across the multiplex
network. Otherwise, it will not affect a finite portion of the nodes
and will die out in a finite time.}
\end{figure}

In this section, by using the SIR model and the bond percolation
theory [7], we propose detailed theoretical analysis that allows us
to accurately calculate the epidemic threshold of multiplex network.
Traditionally the percolation process is parametrized by a
probability $\varphi$, which is the probability that a node is
functioning in the network. In technical terms of percolation
theory, one says that the functional nodes are occupied and
$\varphi$ is called the occupation probability. Through only slight
modifications, the general SIR model can be perfectly mapped into
the bond percolation in complex networks where the spreading rate
corresponds to the probability that a link is occupied in
percolation [7,10]. Therefore, for the multiplex network with two
layers, we can assume that the three types of edges, edge-$A$,
edge-$B$ and edge-$C$ are occupied at the probabilities of
$\lambda_A$,$\lambda_B$ and $\lambda_C$ respectively.

Let $h_A(x)$ ($h_B(x)$, $h_C(x)$) be the generating function [27,28]
for the distribution of the sizes of components which are reached by
an edge with type of edge-$A$ (edge-$B$, edge-$C$) and following it
to one of its ends. Then we have
\begin{equation}\begin{split}
\label{cases}h_A(x) = 1-\lambda_A+
x\lambda_A\times\frac{\sum\limits_{k_M,\ k_A-k_C\geq1}
|k_M|p_{k_M}h_A^{k_A-k_C-1}(x)h_B^{k_B-k_C}
(x)h_C^{k_C}(x)}{\sum\limits_{k_M} |k_M|p_{k_M}},\ \
\end{split}\end{equation}
\begin{equation}\begin{split}
\label{cases}h_B(x) = 1-\lambda_B+
x\lambda_B\times\frac{\sum\limits_{k_M,\ k_B-k_C\geq1}
|k_M|p_{k_M}h_A^{k_A-k_C}(x)h_B^{k_B-k_C-1}
(x)h_M^{k_C}(x)}{\sum\limits_{k_M}
|k_M|p_{k_M}},\end{split}\end{equation}
\begin{equation}\begin{split}
\label{cases}h_C(x) = 1-\lambda_C+
x\lambda_C\times\frac{\sum\limits_{k_M,\ k_C\geq1}
|k_M|p_{k_M}h_A^{k_A-k_C}(x)h_B^{k_B-k_C}
(x)h_M^{k_C-1}(x)}{\sum\limits_{k_M}
|k_M|p_{k_M}},\end{split}\end{equation} where $p_{k_M}$ denotes the
probability that a randomly chosen node of the multiplex network has
the vector degree $k_M$. Later the size of the component formed by
infected nodes will be called the outbreak size.

Generally, an epidemic always starts from a network node, not an
edge, therefore we proceed to analyze the outbreak size distribution
for epidemic sourced from a randomly selected node. If we start at a
randomly chosen node in multiplex network, then we have one such
outbreak size at the end of each edge leaving that node, and hence
the generating function for the outbreak size caused by a network
node is
\begin{equation} \centering{\label{cases}H(x) = x\times\sum\limits_{k_M}
p_{k_M}h_A^{k_A-k_C}(x)h_B^{k_B-k_C} (x)h_C^{k_C}(x).}\end{equation}

Although it is usually impossible to find a closed-form expression
for the complete distribution of outbreak size in a network, we can
find closed-form expressions for the average outbreak size of an
epidemic in multiplex network from Eqs.(6). This average outbreak
size can be derived by taking\ derivates of Eqs.(6) at $x = 1$, and
then we have
\begin{equation}\begin{split}
 \label{cases}<s> =H'(1)=
1+\sum\limits_{k_M} p_{k_M}(k_A-k_C)h'_A(1)+\ \ \ \ \ \ \ \ \ \ \ \
\ \ \ \ \ \ \ \ \ \ \ \ \ \ \ \ \ \\
\sum\limits_{k_M} p_{k_M}(k_B-k_C)h'_B(1)+\sum\limits_{k_M}
p_{k_M}k_Ch'_C(1).\end{split}\end{equation} In Eq.(7), the functions
$h'_A(1)$, $h'_B(1)$ and $h'_C(1)$ can be derived from Eqs.(3)-(5).
Taking derivatives on both sides of Eqs.(3)-(5) at $x = 1$, we have
\begin{equation}
\label{cases}h'_A(1) =
\lambda_A+\lambda_A<k_M>^{-1}(m_{11}h'_A(1)+m_{12}h'_B(1)+m_{13}h'_C(1)),\end{equation}
\begin{equation}
\label{cases}h'_B(1) =  \lambda_B+
\lambda_B<k_M>^{-1}(m_{21}h'_A(1)+m_{22}h'_B(1)+m_{23}h'_C(1)),\end{equation}
\begin{equation}
\label{cases}h'_C(1) =  \lambda_C+
\lambda_C<k_M>^{-1}(m_{31}h'_A(1)+m_{32}h'_B(1)+m_{33}h'_C(1)),\end{equation}
where
\begin{equation*}<k_M>=\sum\limits_{k_M} |k_M|p_{k_M},\end{equation*}
\begin{equation*}m_{11}=\sum\limits_{k_M,k_A-k_C\geq1}|k_M|p_{k_M}(k_A-k_C-1),\end{equation*}
\begin{equation*}m_{12}=\sum\limits_{k_M,k_A-k_C\geq1} |k_M|p_{k_M}(k_B-k_C),\end{equation*}
\begin{equation*}m_{13}=\sum\limits_{k_M,k_A-k_C\geq1} |k_M|p_{k_M}k_C,\end{equation*}
\begin{equation*}m_{21}=\sum\limits_{k_M,k_B-k_C\geq1} |k_M|p_{k_M}(k_A-k_C),\end{equation*}
\begin{equation*}m_{22}=\sum\limits_{k_M,k_B-k_C\geq1}|k_M|p_{k_M}(k_B-k_C-1),\end{equation*}
\begin{equation*}m_{23}=\sum\limits_{k_M,k_B-k_C\geq1} |k_M|p_{k_M}k_C,\end{equation*}
\begin{equation*}m_{31}=\sum\limits_{k_M,k_C\geq1} |k_M|p_{k_M}(k_A-k_C),\end{equation*}
\begin{equation*}m_{32}=\sum\limits_{k_M,k_C\geq1} |k_M|p_{k_M}(k_B-k_C),\end{equation*}
\begin{equation*}m_{33}=\sum\limits_{k_M,k_C\geq1} |k_M|p_{k_M}(k_C-1).\end{equation*}
From Eqs.(8)-(10), we have
\begin{equation}Mh=-<k_M>e,\end{equation} where
$$M=\left (\begin
{array}{ccc}
-\lambda_A^{-1}<k_M>+m_{11} & m_{12} & m_{13}\\
m_{21} & -\lambda_B^{-1}<k_M>+m_{22} & m_{23}\\
m_{31} & m_{32} & -\lambda_C^{-1}<k_M>+m_{33}\\
\end{array} \right ),$$\\
$h=(h'_A(1)\ h'_B(1)\ h'_C(1))^T,$ and $\ e=(1\ 1\ 1)^T$. Therefore,
$h'_A(1)$, $h'_B(1)$, $h'_C(1)$ diverge at the point where
\begin{equation}\emph{\emph{det}}M= 0.\end{equation}
The solution of Eq.(12) yields a set of different critical
2-dimensional points as $\{(\lambda_A, \lambda_B)_c\}$, above any of
which (like the points belonging to the grey shaded area of Fig.2)
$\langle s\rangle$ will diverge, i.e., the epidemic can spread to
the whole network.

Therefore, through above analysis we can see that the epidemic
threshold of the multiplex network with two layers is a set of
2-dimensional points as $\{(\lambda_A, \lambda_B)_c\}$, where
$(\lambda_A, \lambda_B)_c$ satisfies Eq.(12).

\subsection{Calculations of outbreak size}
If a two routes transmitted epidemic with spreading rates
$(\lambda_A, \lambda_B)$ can spread across a multiplex network, the
infected nodes will form into a giant component. Let $u_A$, $u_B$
and $u_C$ be the average probabilities that a node is not connected
to the giant component via edge-$A$, edge-$B$ and edge-$C$,
respectively. According to percolation theory, there are two ways
that may occur: either the edge in question can be unoccupied, or it
is occupied but the node at the other end of the edge is itself not
a member of the giant component. The latter happens only if that
node is not connected to the giant component via any of its other
edges. Thus, we have
\begin{equation}\begin{split}
\label{cases}u_A=1-\lambda_A+ \lambda_A\times
\frac{\sum\limits_{k_M,\ k_A-k_C\geq1}
|k_M|p_{k_M}u_A^{k_A-k_C-1}u_B^{k_B-k_C}
u_C^{k_C}}{\sum\limits_{k_M} |k_M|p_{k_M}},\ \
\end{split}\end{equation}
\begin{equation}\begin{split}
\label{cases}u_B = 1-\lambda_B+ \lambda_B\times
 \frac{\sum\limits_{k_M,\ k_B-k_C\geq1}
|k_M|p_{k_M}u_A^{k_A-k_C}u_B^{k_B-k_C-1}
u_C^{k_C}}{\sum\limits_{k_M} |k_M|p_{k_M}},\end{split}\end{equation}
\begin{equation}\begin{split}
\label{cases}u_C = 1-\lambda_C+ \lambda_C\times
\frac{\sum\limits_{k_M,\ k_C\geq1}
|k_M|p_{k_M}u_A^{k_A-k_C}u_B^{k_B-k_C}
u_C^{k_C-1}}{\sum\limits_{k_M}
|k_M|p_{k_M}}.\end{split}\end{equation} Therefore, the outbreak size
of the two routes transmitted epidemic over the multiplex network
can be calculated by
\begin{equation} \label{cases}s = 1-\sum\limits_{k_M}
p_{k_M}u_A^{k_A-k_C}u_B^{k_B-k_C} u_C^{k_C}.\end{equation}

It is worth to notice that the traditional single route transmitted
epidemic process on single network can be regarded as a special case
of multiplex routes transmitted epidemic process on multiplex
network. When layer-$A$ is regarded as a single network, the
epidemic threshold ${\lambda_A}_c=<k_A>/(<k^2_A>-<k_A>)$ [7] and the
outbreak size $s = 1-\sum\limits_{k_A} p_{k_A}u_A^{k_A}$ [7] can be
obtained from Eq.(12) and Eq.(16) when we let $k_B=0$. That is to
say, the results of this letter are accurate and applicable in a
more general situation.

\section{Simulation results and discussions}
\label{} In this section, we show the theoretical calculation
methods of the epidemic threshold and the outbreak size proposed in
Section 2 is accurate and reasonable by the comparison between the
theoretical values and the experimental results. Three different
types of multiplex network with two layers are constructed where
($i$) both of these two layers are scale-free (SF) networks, denoted
by SF-SF; ($ii$) one layer is Erd\H{o}s-R\'{e}nyi (ER) random
network and the other one is SF network, denoted by ER-SF; and
($iii$) both of these two layers are ER
 random networks, denoted by ER-ER. We use
`X$(a,b)$' to describe a network layer, where `X' refers to the
network type, `$a$' is the network size and `$b$' is the average
degree. For example, SF(2000,3) denotes a SF network comprised of
2000 nodes with average degree 3. Each given simulation result is
averaged over 500 realizations.

\subsection{Epidemic threshold and outbreak size}

In Fig.3(a, b and c), the numerically simulated outbreak sizes of
the two routes transmitted epidemic with spreading rate
$(\lambda_A,\lambda_B)$ are color coded, and the solid black lines
indicate the theoretical epidemic threshold of the multiplex network
calculated according to Eq.(12). We can see that the outbreak size
is a relatively large value, that is, the infected nodes form into a
giant component, when the spreading rate $(\lambda_A,\lambda_B)$ of
the epidemic is a point above the solid black line. Instead, the
outbreak size is a very small value when the spreading rate
$(\lambda_A,\lambda_B)$ is below the solid black line which means
that the theoretical epidemic threshold of the multiplex network
calculated according to Eq.(12) is accurate in judging the epidemic
state. In order to better describe the accurate degree of the
theoretical epidemic threshold, panel d, e and f of Fig.3 show four
longitudinal sections of panel a, b and c, respectively, where
${\lambda_B}_c$ can be regarded as the epidemic threshold of the
multiplex network when $\lambda_A$ are set to some fixed values. It
can also be found that the epidemic could spread across the
multiplex network even if these two layers are well below their
respective epidemic thresholds. The multiplex network
ER(2000,5.922)-ER(2000,5.965) shown in Fig.3(c), is taken as an
example. Since the epidemic threshold of ER network is $1/\langle
k\rangle$ [6] when following the SIR model, where $\langle k\rangle$
is the average degree of the network, we can calculate that the
epidemic thresholds of the layers ER(2000,5.922) and ER(2000,5.965)
are $1/5.922\approx 0.169$ and $1/5.965\approx 0.168$ respectively
when they are regarded as single networks. However, from Fig.3(c) we
can see that the outbreak size of the epidemic is 0.3 if the
spreading rate is $(0.12<0.169, 0.12<0.168)$, and 0.45 if the
spreading rate is $(0.14<0.169, 0.15<0.168)$, and etc. These
illustrate that the epidemic can infect a large number of nodes of
the multiplex network even if these two layers are well below their
respective epidemic thresholds.
\begin{figure}[!htb]
\begin{tabular}{ccc}
\begin{minipage}{0.32\linewidth}
\centering
\includegraphics[scale=0.35,trim=0 0 0 0]{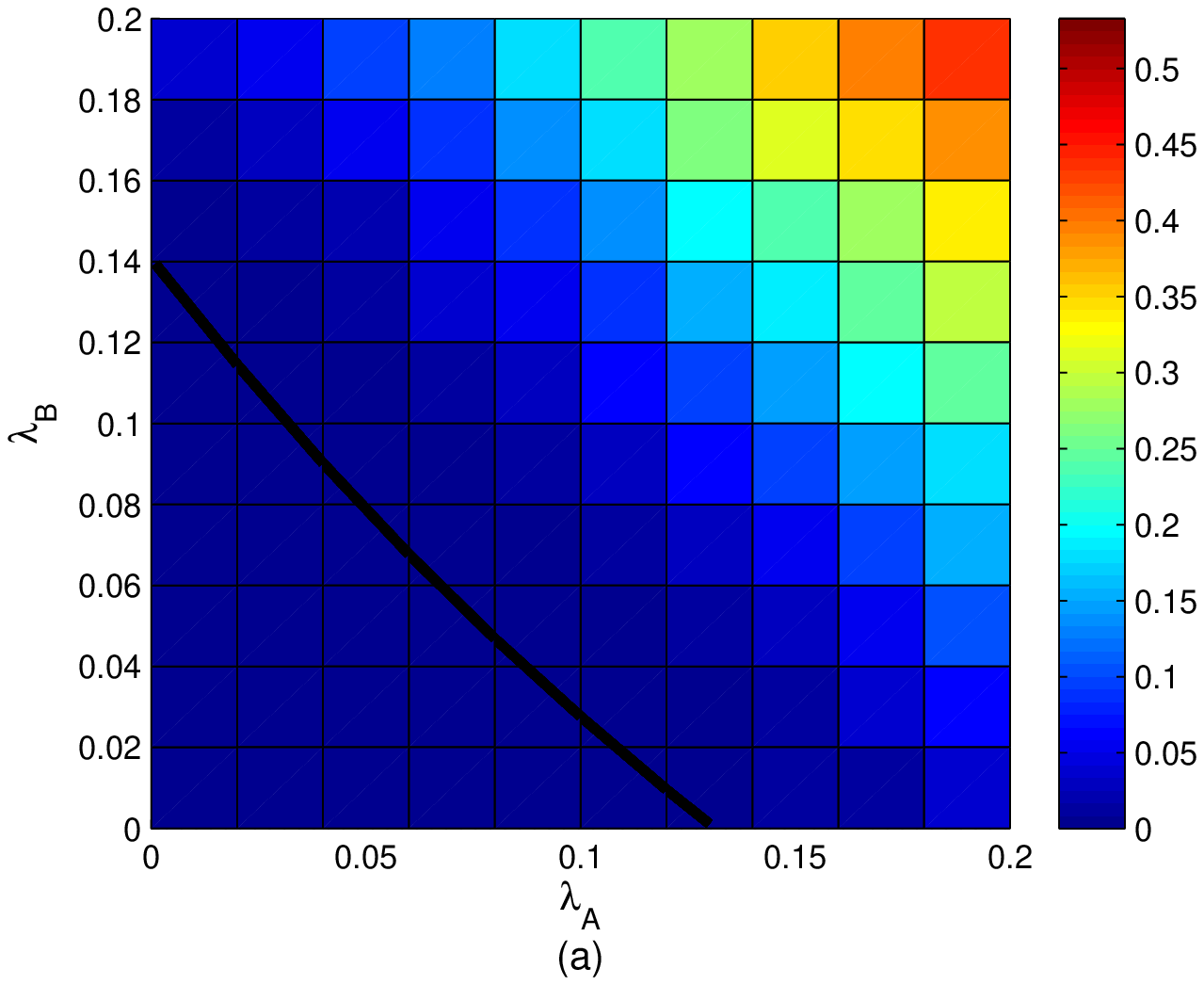}
\end{minipage}
\begin{minipage}{0.32\linewidth}
\centering
\includegraphics[scale=0.35,trim=0 0 0 0]{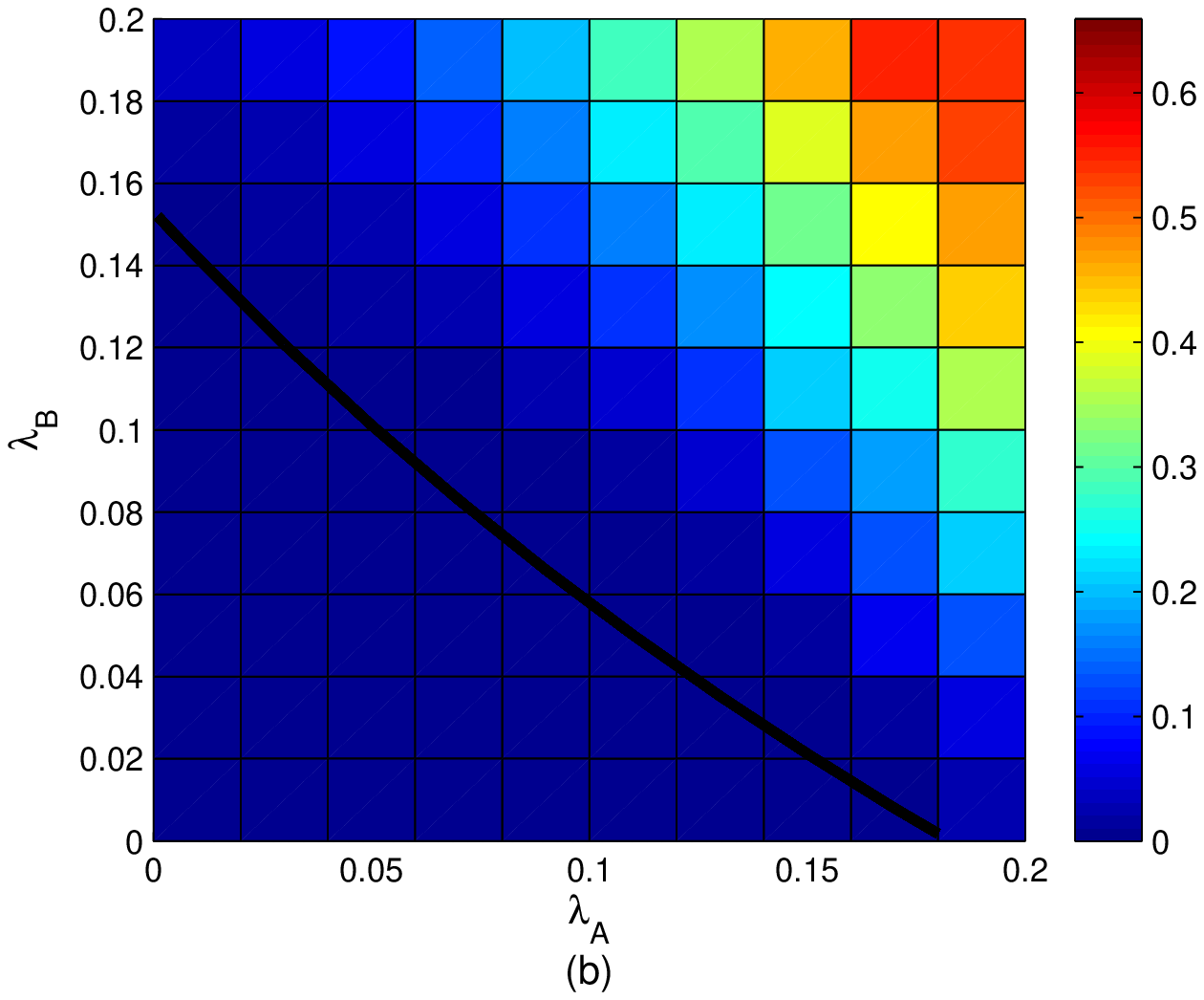}
\end{minipage}
\begin{minipage}{0.32\linewidth}
\centering
\includegraphics[scale=0.35,trim=0 0 0 0]{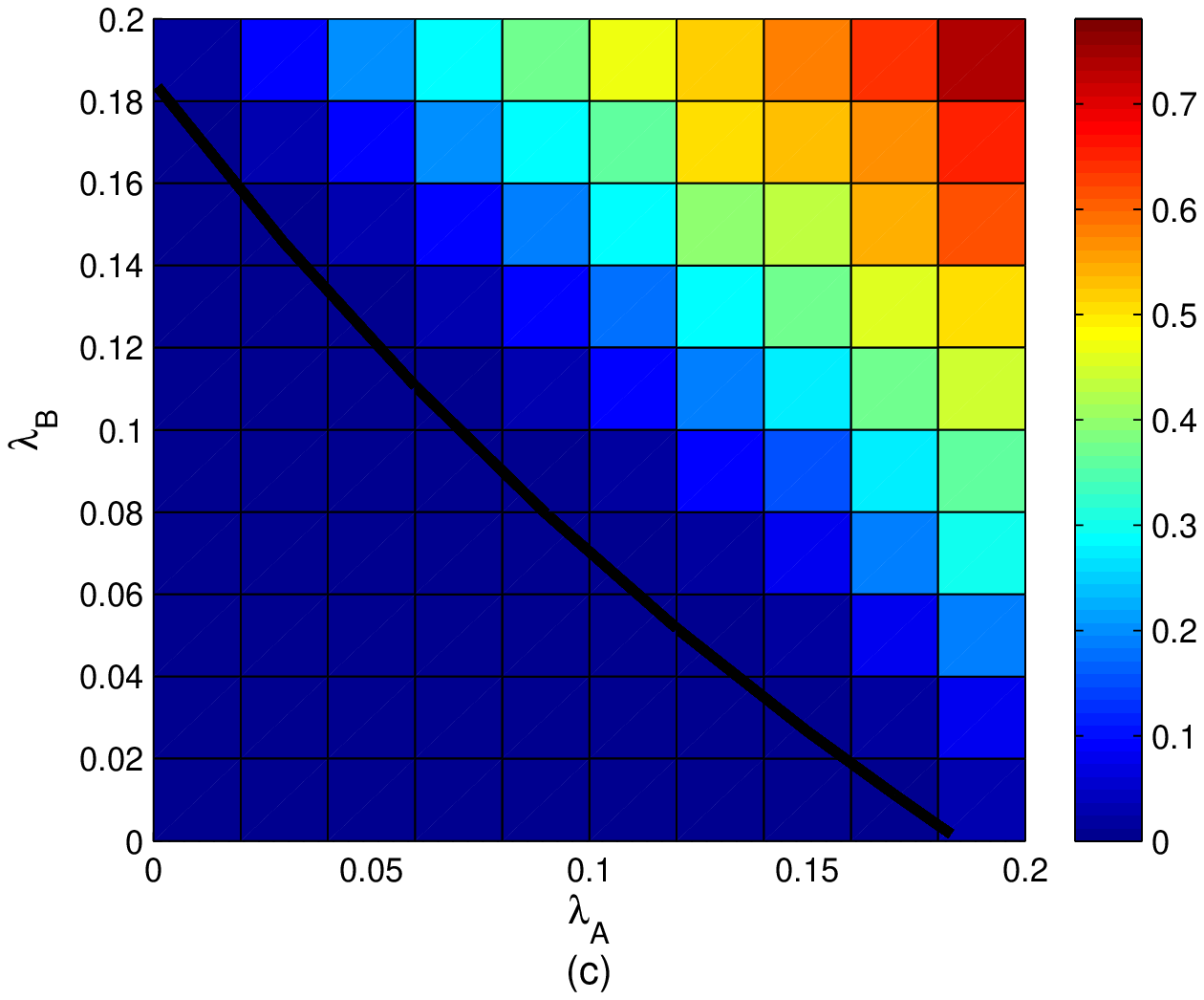}
\end{minipage}\\
\begin{minipage}{0.32\linewidth}
\centering
\includegraphics[scale=0.35,trim=0 0 0 0]{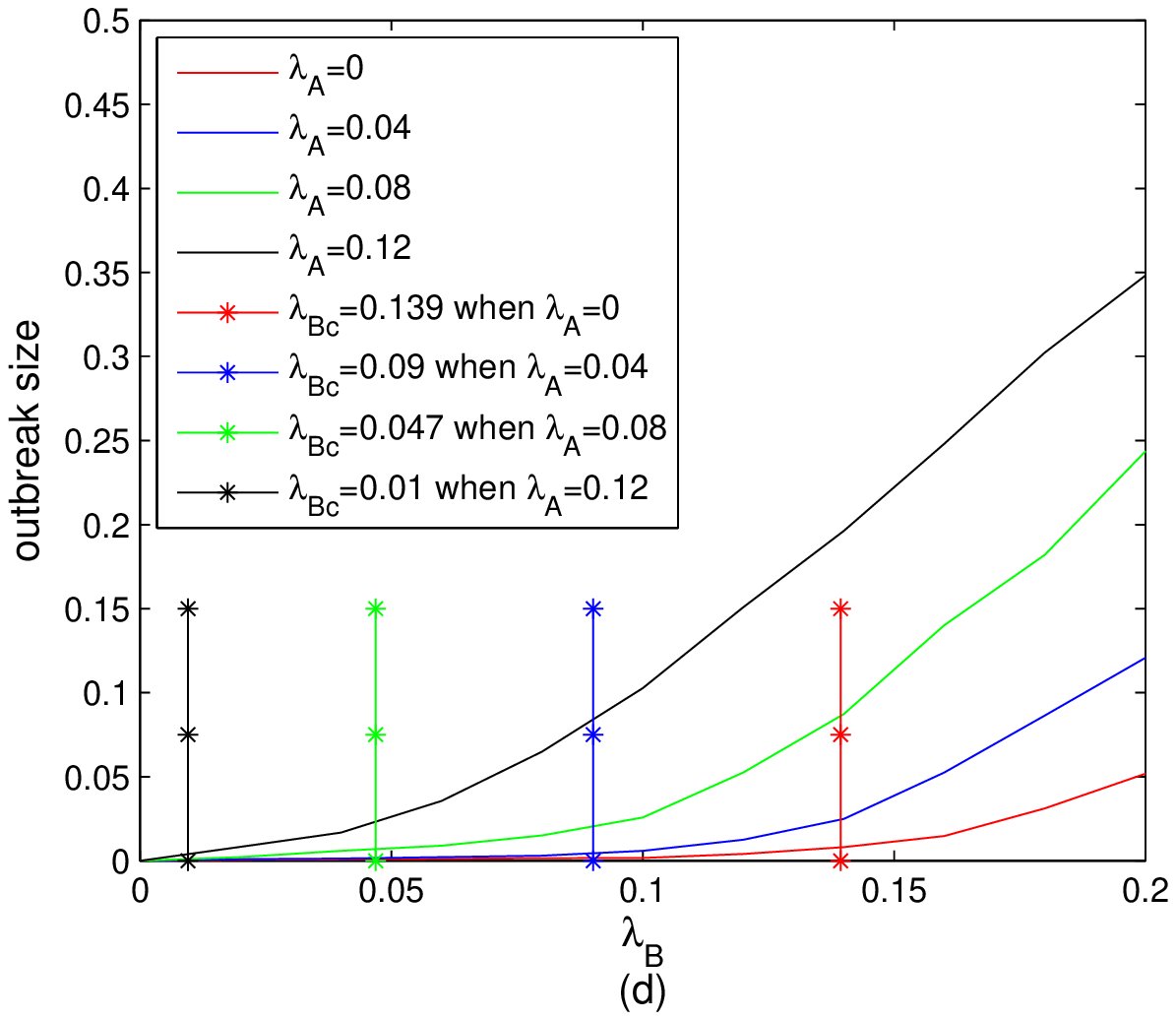}
\end{minipage}
\begin{minipage}{0.32\linewidth}
\centering
\includegraphics[scale=0.35,trim=0 0 0 0]{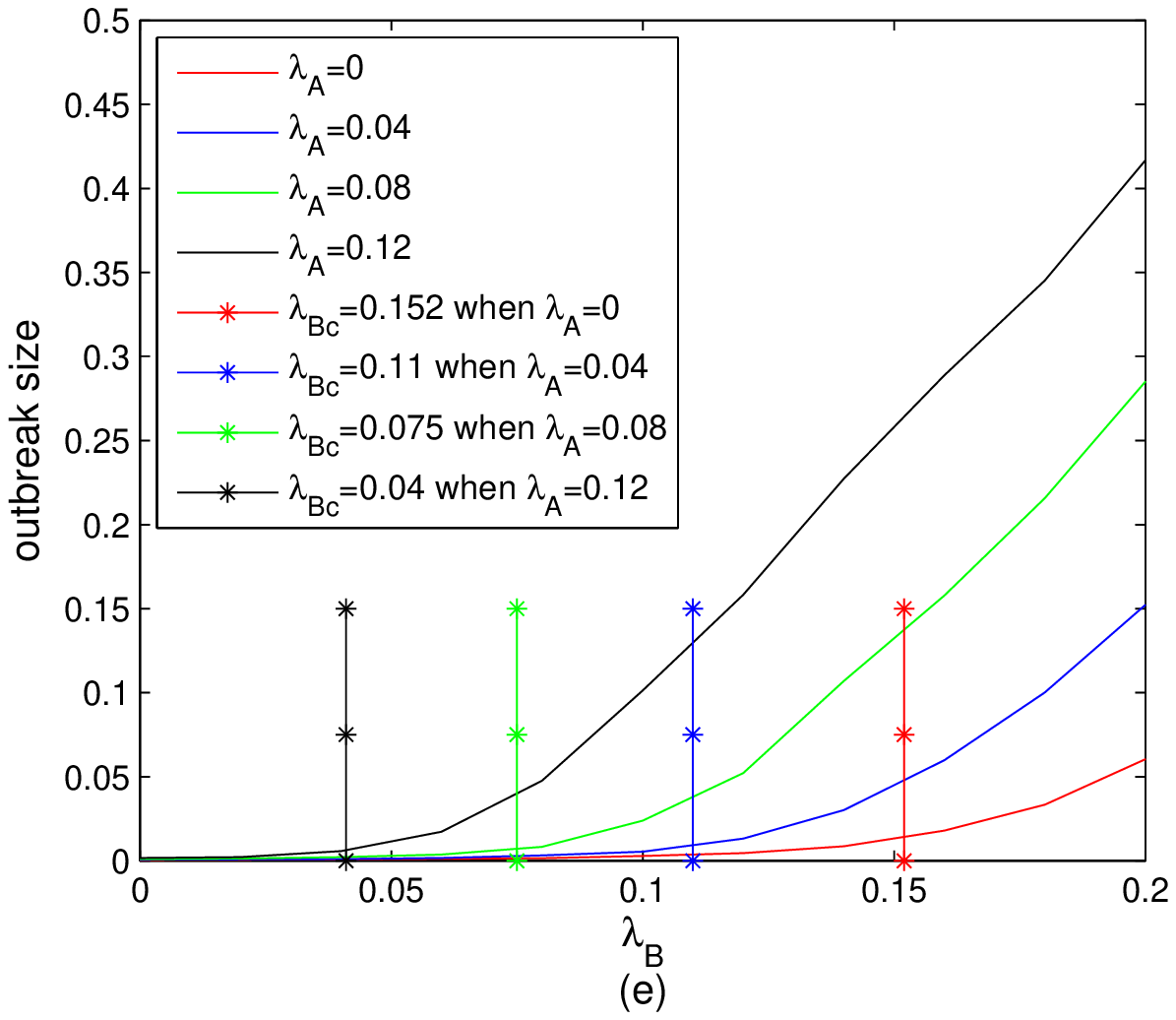}
\end{minipage}
\begin{minipage}{0.32\linewidth}
\centering
\includegraphics[scale=0.35,trim=0 0 0 0]{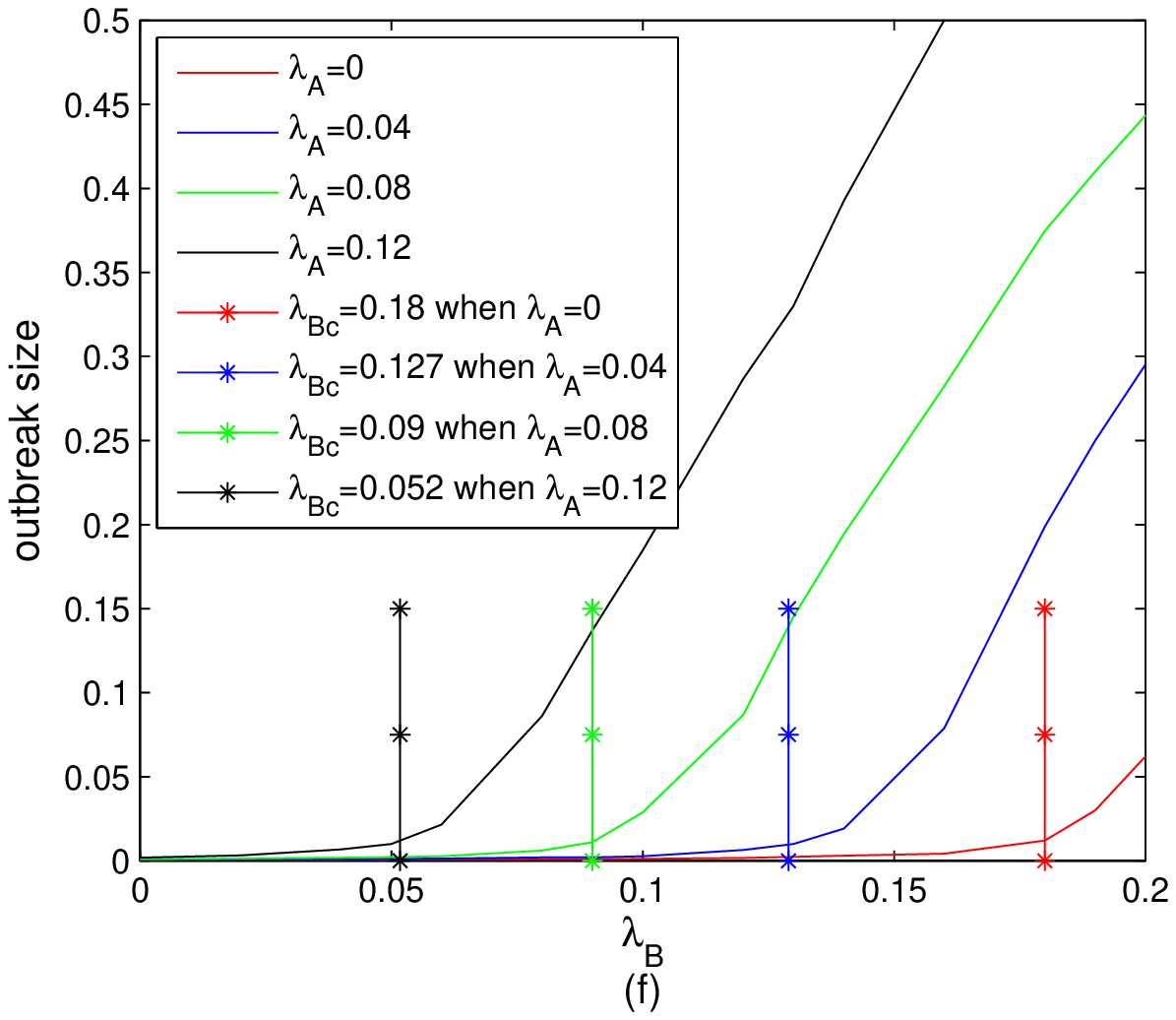}
\end{minipage}
\end{tabular}
\caption{(Color online) Panel a, b and c show theoretical epidemic
threshold of the multiplex network calculated according to Eq.(12)
(solid black line) and the numerically simulated outbreak sizes of
the epidemic with spreading rate $(\lambda_A,\lambda_B)$ (color
coded). Panel d, e and f show four longitudinal sections of panel a,
b and c, respectively. Three multiplex networks are (a,d)
SF(2000,3.997)-SF(2000,3.998), (b,e) ER(2000,5.883)-SF(2000,3.997)
and (c,f) ER(2000,5.922)-ER(2000,5.965), respectively.}
\end{figure}

Fig.4 shows the numerically simulated and the theoretical outbreak
sizes of the epidemic as a function of $\lambda_B$ when $\lambda_A$
are set to some fixed values. The theoretical results are calculated
according to Eq.(16). From Fig.4 we can see that theoretical results
are in good agreement with the experimental results.
\begin{figure}[!htb]
\begin{tabular}{ccc}
\begin{minipage}{0.32\linewidth}
\centering
\includegraphics[scale=0.34,trim=0 0 0 0]{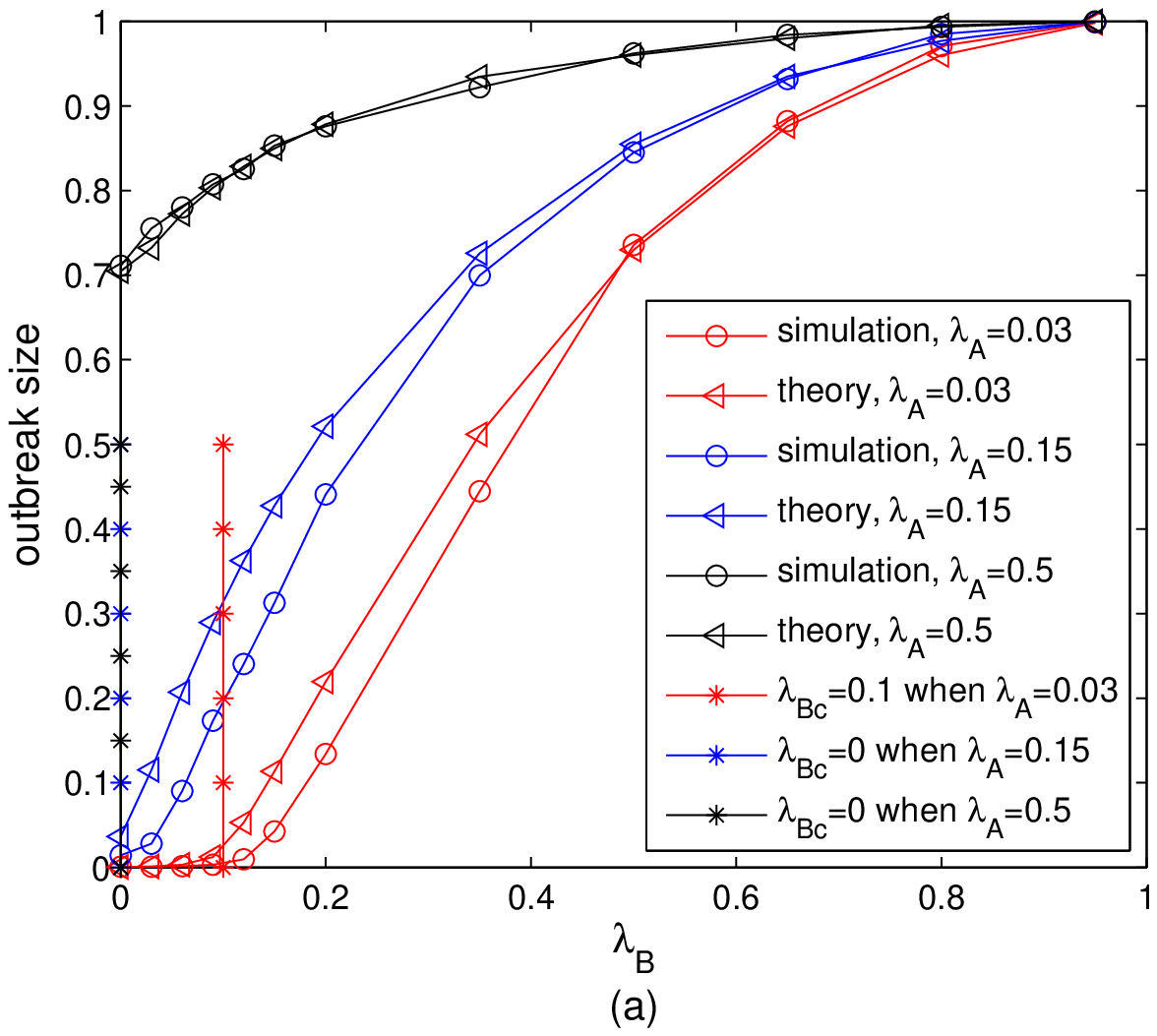}
\end{minipage}
\begin{minipage}{0.32\linewidth}
\centering
\includegraphics[scale=0.34,trim=0 0 0 0]{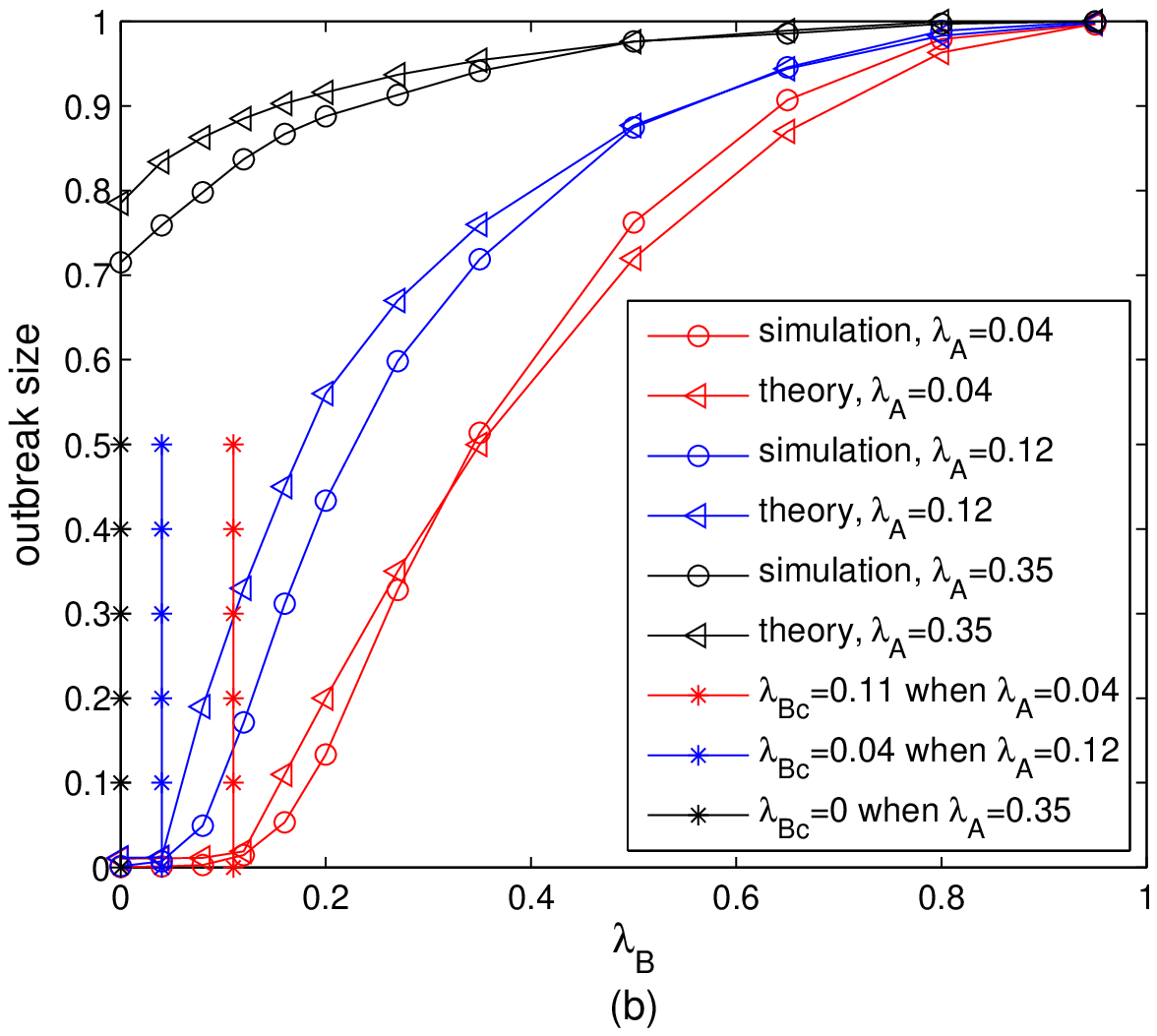}
\end{minipage}
\begin{minipage}{0.32\linewidth}
\centering
\includegraphics[scale=0.34,trim=0 0 0 0]{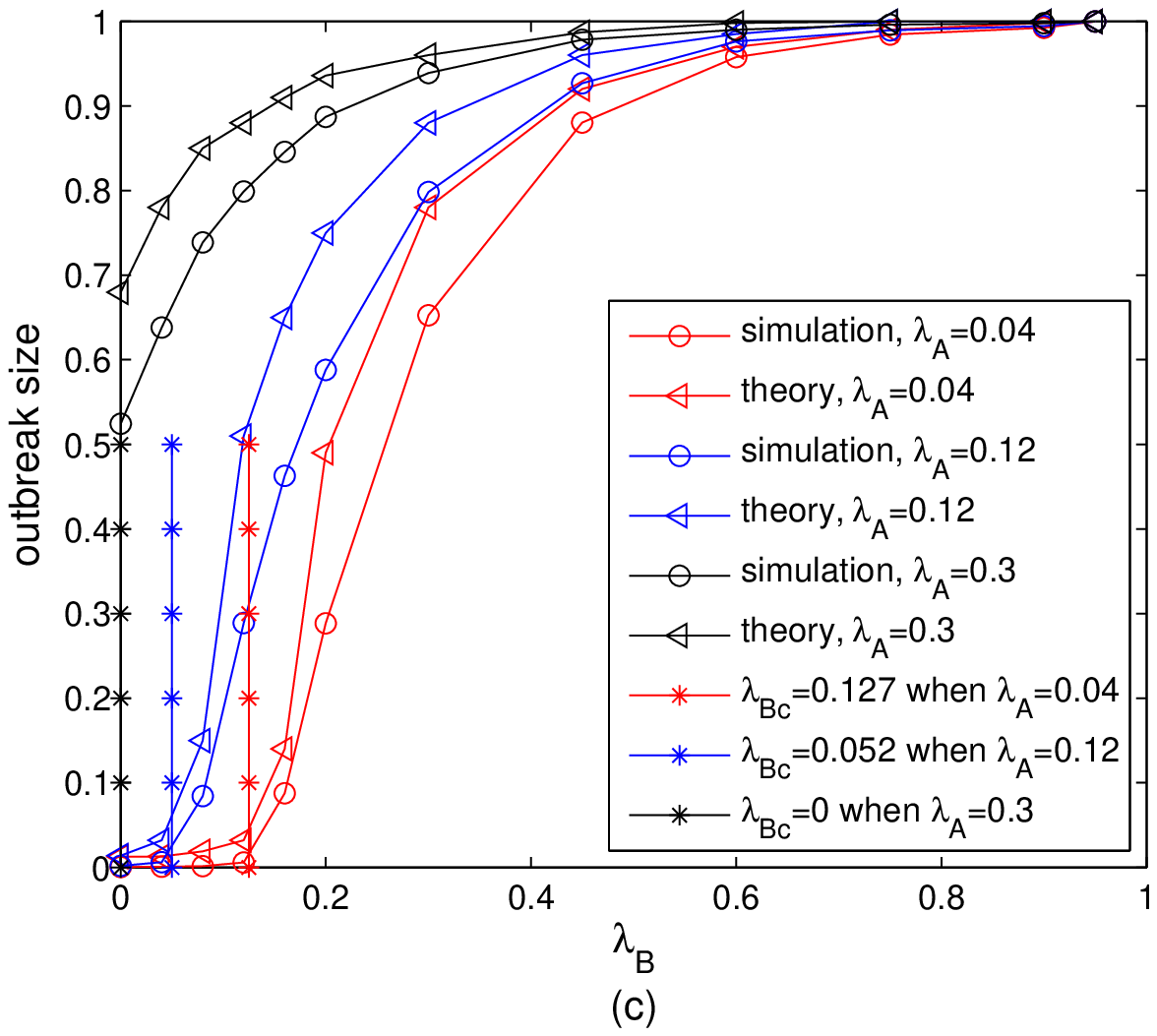}
\end{minipage}
\end{tabular}
\caption{(Color online) Numerically simulated and theoretical
outbreak sizes of the epidemic as a function of $\lambda_B$ when
$\lambda_A$ are set to some fixed values. Three multiplex networks
are (a) SF(2000,3.997)-SF(2000,3.998), (b)
ER(2000,5.883)-SF(2000,3.997) and (c) ER(2000,5.922)-ER(2000,5.965),
respectively.}
\end{figure}

\subsection{ASN and DDC}

In the real world, the nodes of the multiplex network may have some
same neighbors in the two layers. In Section 2, we used a measure
ASN to assess the average similarity of the neighbors from different
layers of nodes in the multiplex network. Another quantity DDC is
also developed to describe the correlation of nodes' degrees in one
layer and that in another layer. While the topologies of the two
layers remain unchanged, the topology of the multiplex network could
be affected to some extent by the ASN and DDC, and may ultimately
impact the processes of the epidemic over it.

It is easy to achieve any targeted value of ASN for SF-SF and ER-ER.
These two ER(SF) network layers can be obtained by
randomly(preferentially) adding edges to a same ER(SF) network which
has been constructed, respectively. The value of ASN is determined
by the number of the edges added and the edges of the initial
network. It is hard, however, to achieve a large range of ASN for
ER-SF. Assume that the nodes have same tabs in each layer, then we
can get some different values of ASN of ER-SF by randomly exchanging
the tabs of nodes for one layer. In the ER-SF model shown in
Figs.5(b,e), the value of ASN roughly lies in the interval [0.02
0.14], while for the SF-SF and ER-ER models, the corresponding
intervals are all [0, 1].

In the following experiments, we assume the epidemic has the same
spreading rates when propagates on the two layers, i.e.,
$\lambda_A=\lambda_B$. As shown in Fig.5, the epidemic threshold and
the outbreak size are seldom affected by the ASN no matter what type
of the multiplex network. This can be understood that when the
topologies of the two layers remain unchanged, high ASN means nodes
can affect much of their neighbors with the large spreading rate
$\lambda_C$, but the average number of their neighbors is relatively
small. Instead, although low ASN implies the nodes have more
neighbors, most of the spreading rates between them and their
neighbors are the relatively small $\lambda_A$ and $\lambda_B$. In
such cases, the average number of new infected nodes at a time step
may be equivalent no matter what the values of ASN.
\begin{figure}[!htb]
\begin{tabular}{ccc}
\begin{minipage}{0.32\linewidth}
\centering
\includegraphics[totalheight=8.5cm]{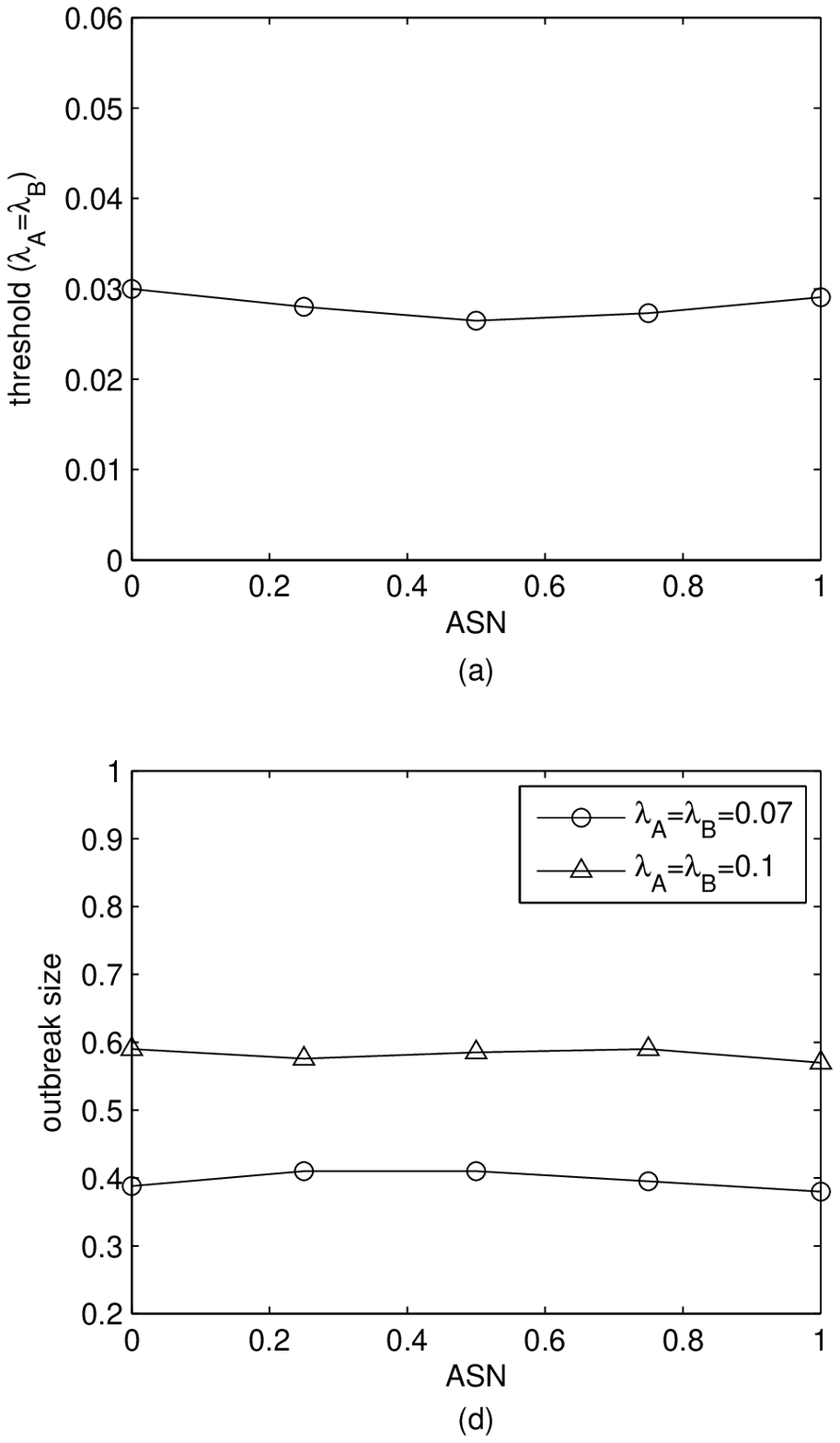}
\end{minipage}
\begin{minipage}{0.32\linewidth}
\centering
\includegraphics[totalheight=8.5cm]{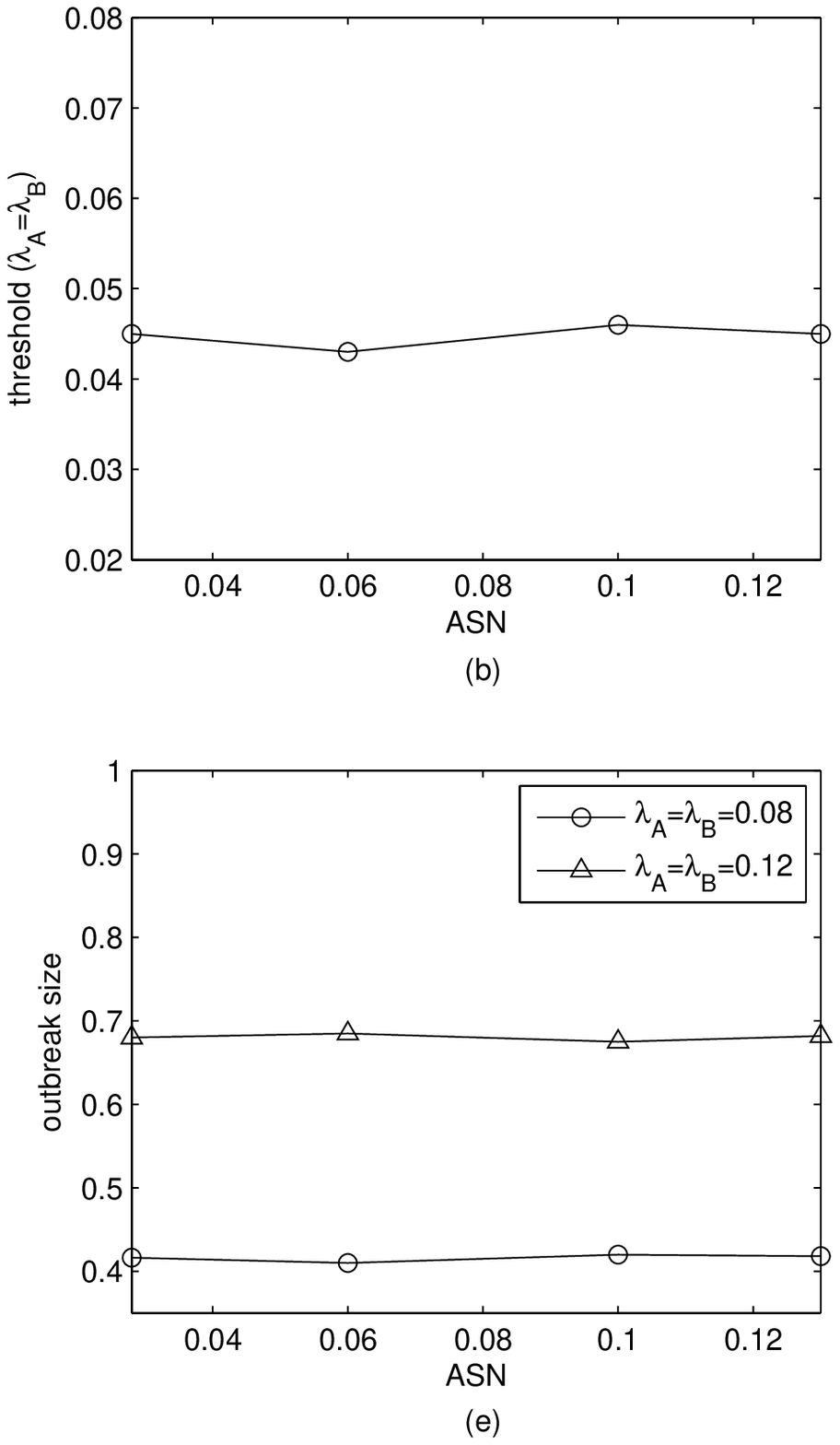}
\end{minipage}
\begin{minipage}{0.32\linewidth}
\centering
\includegraphics[totalheight=8.5cm]{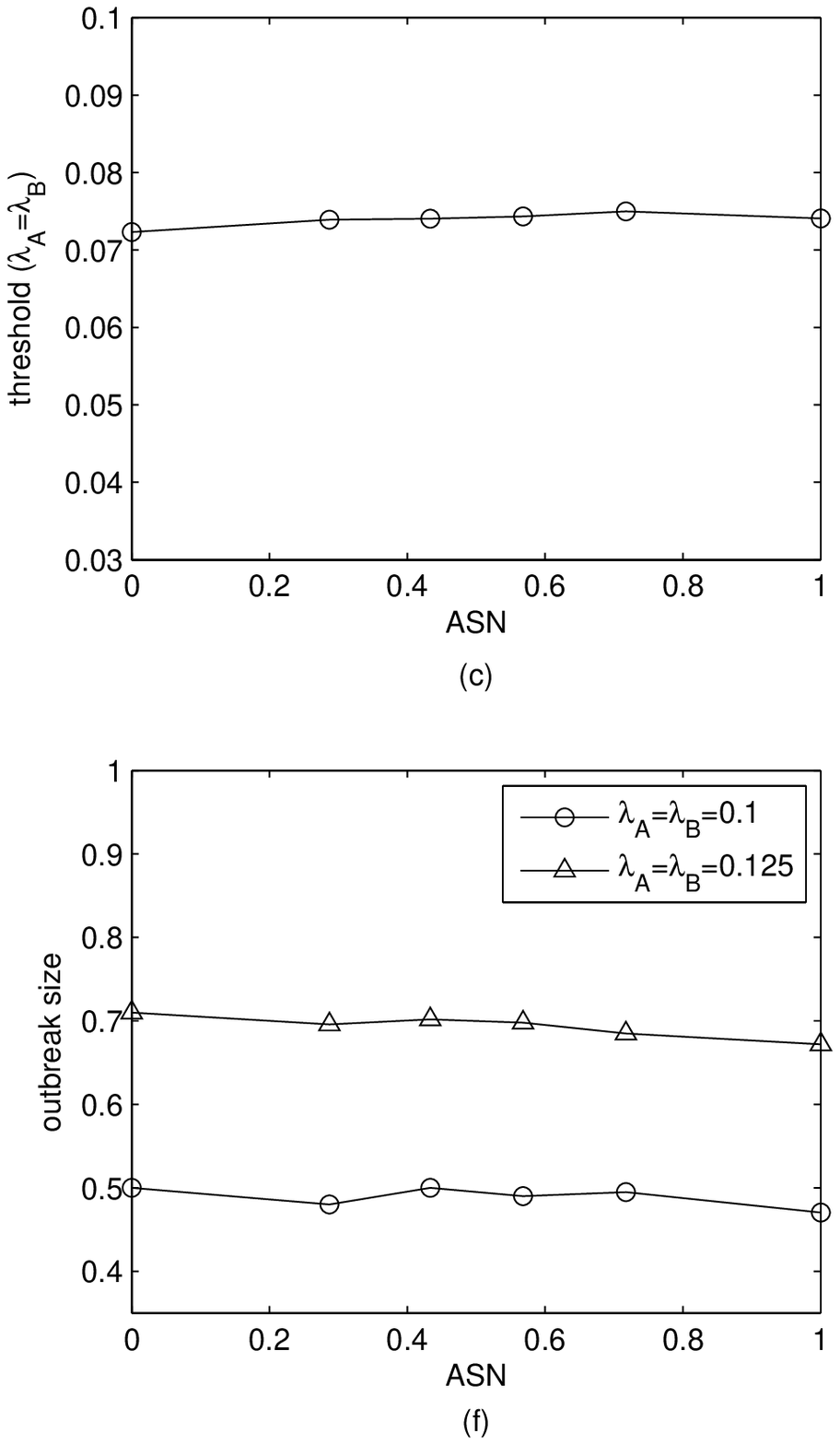}
\end{minipage}
\end{tabular}
\caption{Epidemic threshold and outbreak size for epidemic spreading
on (a,d) SF-SF; (b,e) ER-SF; and (c,f) ER-ER.}
\end{figure}

Fig.6 shows the influences of DDC on the epidemic threshold and the
outbreak size. Different values of DDC can be achieved by randomly
exchanging the tabs of nodes of one layer. As shown in Fig.6, a
higher DDC can lead to a much lower epidemic threshold  and a
relatively smaller outbreak size no matter what type of the
multiplex network. The reasons can be explained as follows: high DDC
means that high degree nodes in one layer are also high nodes in
another layer and low degree nodes in one layer also low degree
nodes in another layer, which leads to increasing differences
between the degrees of nodes in the multiplex network. Instead, low
DDC leads to decreased differences between the degrees of nodes in
the multiplex network. That is, high DDC makes the SF-SF and ER-SF
be the strengthened inhomogeneous networks and ER-ER a proximate
inhomogeneous network, low DDC however makes the three types of
multiplex network be the proximate homogeneous networks. It is known
that [29], the epidemic in the inhomogeneous network has a faster
spread since the existence of high degree nodes and smaller outbreak
size since the low degree nodes are not prone to be infected, than
in the homogeneous network when these two networks have the same
average degrees. This theory perfectly explains the results of the
experiments.
\begin{figure}[!htb]
\begin{tabular}{ccc}
\begin{minipage}{0.32\linewidth}
\centering
\includegraphics[totalheight=8.5cm]{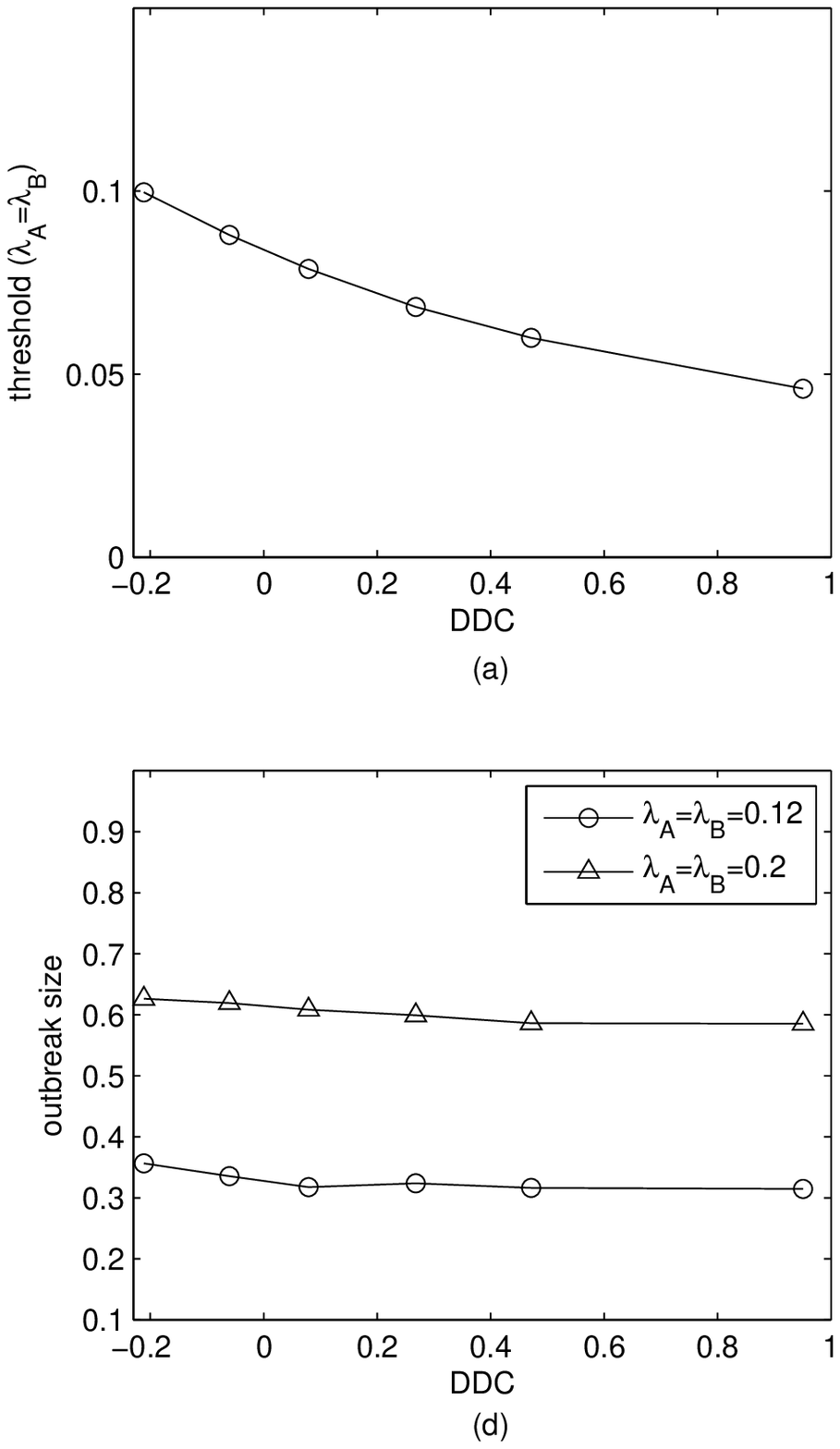}
\end{minipage}
\begin{minipage}{0.32\linewidth}
\centering
\includegraphics[totalheight=8.5cm]{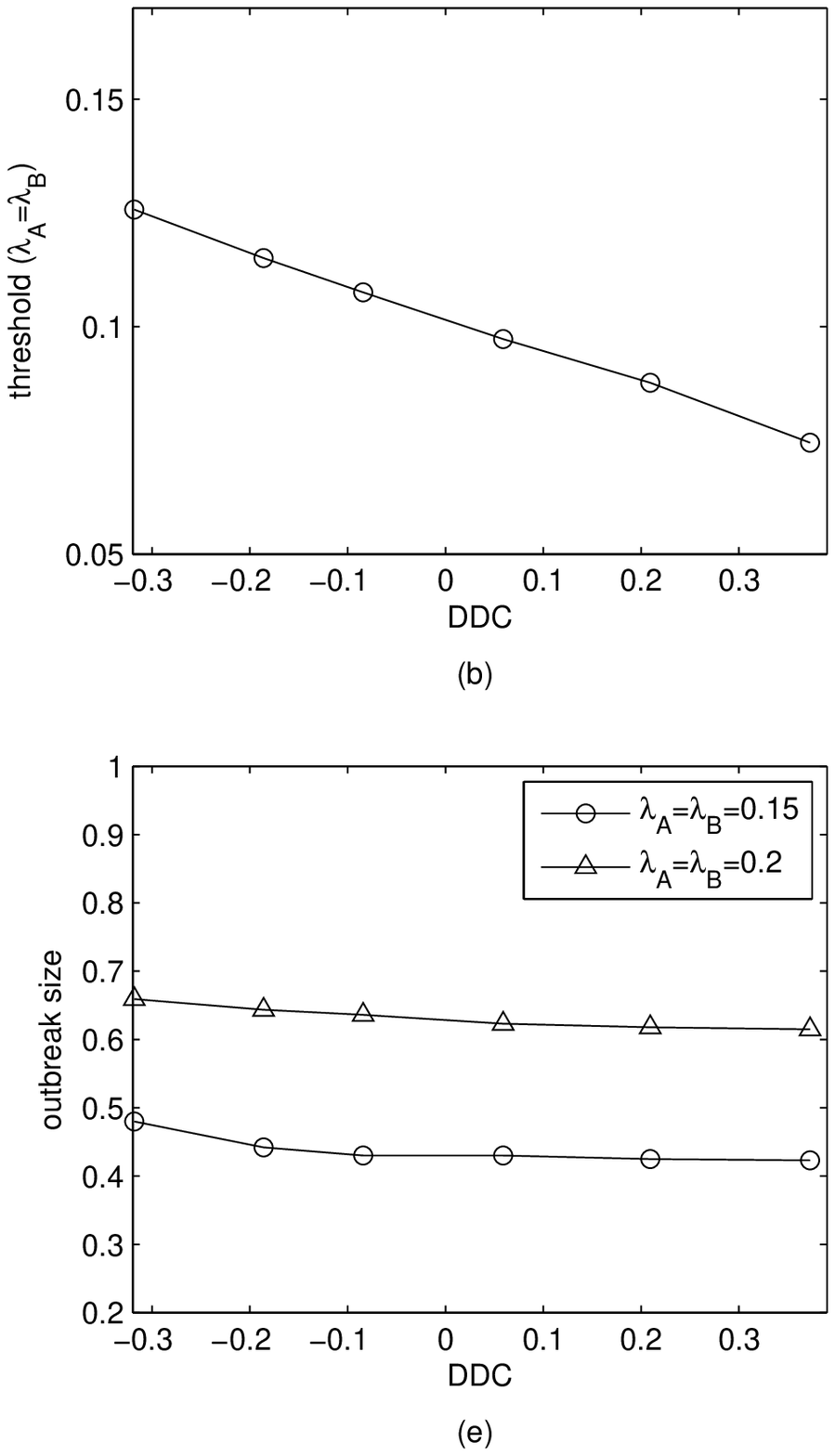}
\end{minipage}
\begin{minipage}{0.32\linewidth}
\centering
\includegraphics[totalheight=8.5cm]{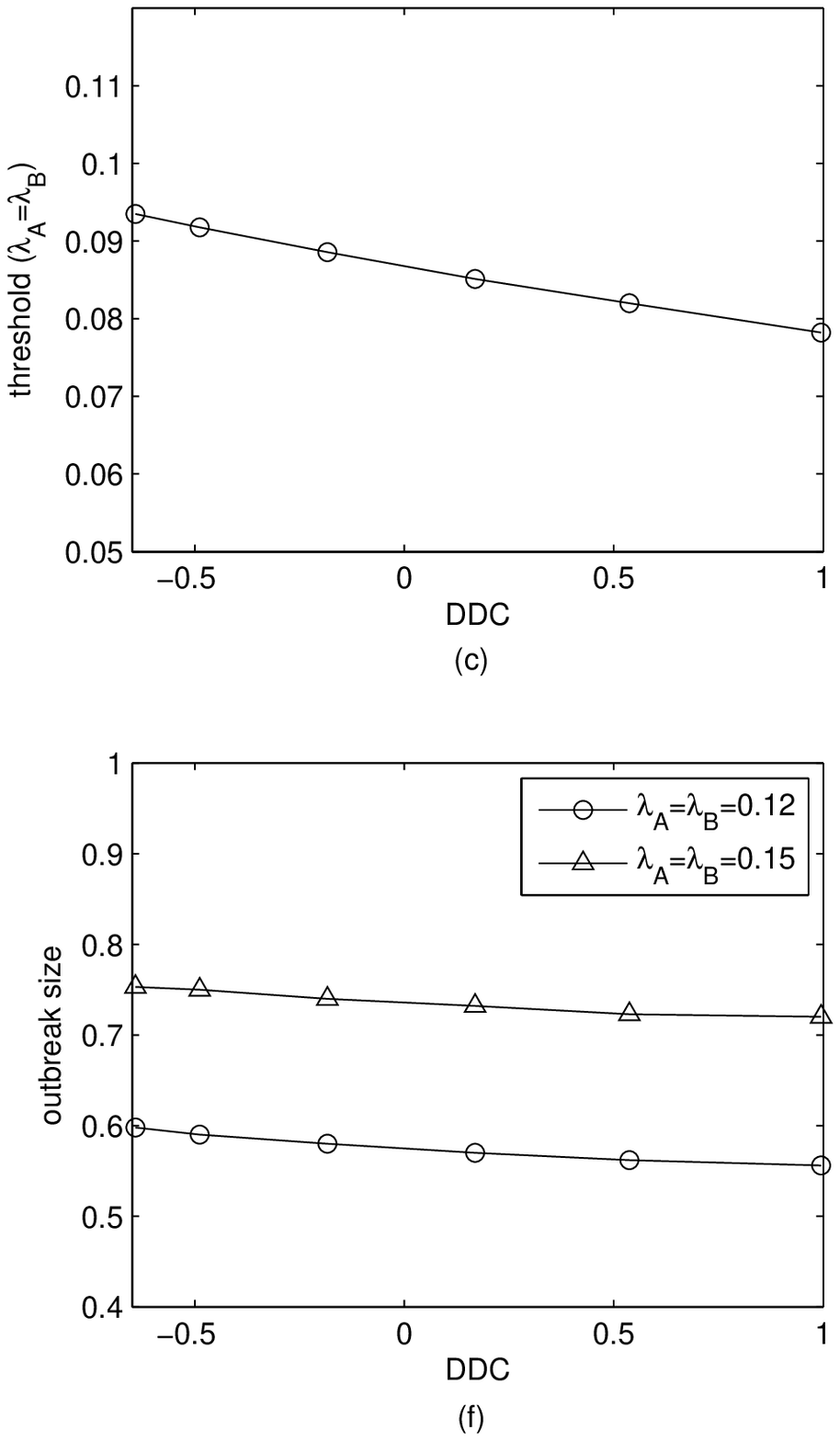}
\end{minipage}
\end{tabular}
\caption{Epidemic threshold and outbreak size for epidemic spreading
on (a,d) SF(2000,3.997)-SF(2000,3.995); (b,e)
ER(2000,4.005)-SF(2000,3.997); and (c,f)
ER(2000,5.950)-ER(2000,5.956).}
\end{figure}

\section{Conclusions}
\label{} In this letter, we demonstrated the dynamics of two routes
transmitted epidemic spreading on multiplex network with two network
layers following the SIR model. Our main contributions can be
summarized as follows: (1) We presented the multiple routes
transmitted system of epidemics and derived equations to accurately
calculate the epidemic threshold and the outbreak size in the
multiplex network. (2) We found that the epidemics could spread
across the multiplex network even if the two layers are well below
their respective epidemic thresholds. (3) We proposed two quantities
for measuring the level of inter-similarity between two layers. ASN
evaluates how many neighbors of nodes in one layer are also their
neighbors in another layer which is found barely affect the epidemic
threshold and the outbreak size. DDC describes the correlation of
node's degree in one layer and that in another layer. It is found
that higher DDC could lead to much lower epidemic threshold and
relatively smaller outbreak size.

Although we only consider the two routes transmitted epidemic
process on multiplex network with two network layers, the proposed
research methods are easily extended to analyze the epidemics which
spread via any number of transmission routes. Our research provides
useful tools and novel insights for further studies of dynamics of
multiple routes transmitted epidemic spreading on the multiplex
networks.

\section{Acknowledgement}
This paper was supported by the Foundation for the Author of
National Excellent Doctoral Dissertation of PR China (Grant No.
200951), the National Natural Science Foundation of China (Grant
Nos. 61202362, 61170269, 61121061), the Asia Foresight Program under
NSFC Grant (Grant No. 61161140320).

\end{document}